\documentclass[prl,twocolumn,showpacs,superscriptaddress,preprintnumbers]{revtex4}

\usepackage{graphicx}
\usepackage{dcolumn}
\usepackage{amsmath}
\usepackage{color}

\begin{document}

\title{Emergence of quantum critical charge and spin-state fluctuations near the  pressure-induced Mott transition in MnO, FeO, CoO, and NiO}

\author{I. Leonov}
\affiliation{Institute of Metal Physics, Sofia Kovalevskaya Street 18, 620219 Yekaterinburg GSP-170, Russia}
\affiliation{Department of theoretical physics and applied mathematics, Ural Federal University, Mira St. 19, 620002 Yekaterinburg, Russia}
\affiliation{Materials Modeling and Development Laboratory, National University of Science and Technology 'MISIS', 119049 Moscow, Russia}

\author{A. O. Shorikov}
\affiliation{Institute of Metal Physics, Sofia Kovalevskaya Street 18, 620219 Yekaterinburg GSP-170, Russia}
\affiliation{Department of theoretical physics and applied mathematics, Ural Federal University, Mira St. 19, 620002 Yekaterinburg, Russia}

\author{V. I. Anisimov}
\affiliation{Institute of Metal Physics, Sofia Kovalevskaya Street 18, 620219 Yekaterinburg GSP-170, Russia}
\affiliation{Department of theoretical physics and applied mathematics, Ural Federal University, Mira St. 19, 620002 Yekaterinburg, Russia}

\author{I. A. Abrikosov}
\affiliation{Department of Physics, Chemistry and Biology (IFM), Link\"oping University, SE-58183 Link\"oping, Sweden}
\affiliation{Materials Modeling and Development Laboratory, National University of Science and Technology 'MISIS', 119049 Moscow, Russia}

\date{\today}

\begin{abstract}

We perform a comprehensive theoretical study of the pressure-induced evolution of the electronic structure, magnetic state, and phase stability of the late transition metal monoxides MnO, FeO, CoO, and NiO using a fully charge self-consistent DFT+dynamical mean-field theory method. 
Our results reveal that the pressure-induced Mott insulator-to-metal phase transition in MnO-NiO is accompanied by a simultaneous collapse of local magnetic moments and lattice volume, implying a complex interplay between chemical bonding and electronic correlations.
We compute the pressure-induced evolution of relative weights of the different valence states and spin-state configurations. Employing the concept of fluctuating valence in a correlated solid, we demonstrate that in MnO, FeO, and CoO a Mott insulator-metal transition and collapse of the local moments is accompanied by a sharp crossover of the spin-state and valence configurations. Our microscopic explanation of the magnetic collapse differs from the accepted picture and points out a remarkable dynamical coexistence (frustration) of the high-, intermediate-, and low-spin states. In particular, in MnO, the magnetic collapse is found to be driven by the appearance of the intermediate-spin state (IS), competing with the low-spin (LS) state; in FeO, we observe a conventional high-spin to low-spin (HS-LS) crossover. Most interestingly, in CoO, we obtain a remarkable (dynamical) coexistence of the HS and LS states, i.e., a HS-LS frustration, up to high pressure. Our results demonstrate the importance of quantum fluctuations of the valence and spin states for the understanding of quantum criticality of the Mott transitions.

\end{abstract}


\maketitle

\section*{I. INTRODUCTION}

The Mott metal-insulator transition caused by the mutual interaction between electrons is one of the most fundamental concepts of condensed matter physics \cite{Review}. This phenomenon occurs in Mott insulators, e.g., under pressure or doping of charge carries and has attracted much interest in view of its importance for unconventional high-$T_c$ superconductivity in cuprates and iron-based materials, as well as for the understanding of manganites, showing colossal magnetoresistance, and heavy-fermion behavior in the $f$-electron systems \cite{Review,hightc}. Even today, it remains among the main highly-debated topics of condensed-matter physics \cite{Review,hightc,Earth}.

The series of transition metal monoxides, MnO, FeO, CoO, and NiO, containing the partially filled $3d$ shell (with an electronic configuration ranging from $3d^5$ to $3d^8$, respectively), are perhaps among the most extensively studied examples of a Mott transition. At high temperature, these materials are known to exhibit a pressure-induced Mott transition in their paramagnetic phase with a cubic rocksalt B1 crystal structure \cite{Review,CM93,HS-LS+MIT}. 
Below the N\'eel temperatures, ranging from $T_N\sim 116$ K to 523 K for MnO to NiO, respectively, these materials undergo a structural phase transition into a distorted rhombohedral phase \cite{HS-LS+MIT}.
The Mott transition is of first-order, often accompanied by a dramatic reduction of the unit-cell volume, implying a coupling between electronic and lattice degrees of freedom. In MnO, FeO, and CoO it is followed by a magnetic collapse -- a remarkable reduction of the local magnetic moments of transition metal ions \cite{CM93,HS-LS+MIT}.
Moreover, MnO, FeO, and CoO exhibit rich allotropic behavior at high-pressures revealing a complex interplay between electron correlation and delocalization (i.e., metallic character) along with
changes in crystal structure and transition metal spin-state \cite{HS-LS+MIT,Zhang,Greenberg}.
In spite of intensive research that arguably provided quite complete understanding of the Mott transition \cite{Review}, there are still many electronic and magnetic phenomena near the Mott insulator-to-metal transition (IMT) which are not fully understood from a theoretical point of view, especially in the high-pressure and high-temperature regime \cite{HS-LS+MIT}. These are, for example, the nature of quantum criticality of the Mott transition, appearance of strange metal and non-Fermi liquid behaviors in proximity to the Mott IMT, which are actively debated in the literature \cite{q_criticality_teor, q_criticality_exp}.

In practice, many of the electronic, magnetic, and structural properties of real materials can be explained using, e.g., band-structure methods \cite{CM93,band_structure,DLM_Zunger,DLM}. 
While these techniques often provide a quantitative description of the static electronic properties of correlated systems, such as an energy gap and magnetic moments, band-structure methods neglect electronic dynamics.
As a result, these methods cannot capture all the generic aspects of a Mott IMT, such as a formation of the lower- and upper-Hubbard (incoherent) subbands, to explain coherence-incoherence crossover, quasiparticle behavior and strong renormalization of the electron mass in the vicinity of a Mott IMT, all because of the neglecting of the effect of strong correlations \cite{Review}.
Here, we overcome this obstacle by using a DFT+DMFT approach~\cite{DMFT,LDA+DMFT} (DFT+DMFT: density functional plus dynamical mean-field theory) which merges \emph{ab initio} band-structure methods and dynamical mean-field theory of correlated electrons \cite{DMFT}, 
providing a good quantitative description of the electronic and structural properties of strongly correlated systems \cite{SK01,MH03,KH04,KK09,LB08,LP11,GP13,Fe_based_SC,PM14,KL08,IL15,BK12,PhysRevB.100.245109,FeS,FeO2}. In particular, this advanced theory makes it possible to determine the electronic structure, magnetic state, and lattice stability of paramagnetic correlated materials at finite temperatures, e.g., near the Mott IMT \cite{KH04,KK09,LB08,PM14,KL08,IL15,BK12,PhysRevB.100.245109,FeS}.


In this paper, we study the pressure-induced evolution of electronic and magnetic properties of the late transition metal monoxides, from MnO to NiO using a state-of-the-art self-consistent over charge density DFT+DMFT method \cite{charge-sc-LDA+DMFT} implemented with plane-wave pseudopotentials \cite{Pseudo}. We explore the evolution of their electronic structure and magnetic states near the pressure-induced Mott IMT, which was shown to be accompanied by a magnetic collapse -- a transformation from the high-spin to low-spin state (HS-LS), all in the B1 crystal structure \cite{CM93,IL15}.
Here, we focus on their high-temperature properties in the paramagnetic state well above the N\'eel temperature to exclude the complications associate with a structural phase transformation, e.g., in a low-temperature distorted rhombohedral phase.
We obtain that under pressure MnO-NiO exhibit a Mott IMT which is accompanied by a simultaneous collapse of local magnetic moments and lattice volume, with a transition pressure $p_c$ varying from $\sim$145 to 40 GPa, upon moving from MnO to CoO, and $p_c \simeq 429$ GPa for NiO. 
We show that in MnO, FeO, and CoO the Mott IMT and the concomitant collapse of the local moments is accompanied by a sharp crossover of the valence state, implying the importance of the valence fluctuations for understanding their electronic states in the vicinity of the pressure-driven IMT. 
We give a novel microscopic explanation of the magnetic collapse of these compounds, revealing a remarkable quantum superposition of the high-, intermediate-, and low-spin states near the Mott transition, i.e., a HS-LS and IS frustration.
Our results provide a novel microscopic explanation of the magnetic collapse of all these compounds. In fact, in MnO the magnetic collapse is found to be driven by the appearance of the intermediate-spin state (IS), strongly competing with the LS state; in FeO we observe a conventional HS-LS crossover. Most interestingly, in CoO we obtain a remarkable coexistence (frustration) of the HS and LS states, up to high compression.
Overall, our results qualitatively improve understanding of the pressure-induced evolution of the electronic  and magnetic structure in correlated insulators, which may have important implications for the theoretical picture of quantum criticality of the Mott transitions \cite{q_criticality_exp}.

\section*{II. METHOD}

We employ the DFT+DMFT approach \cite{charge-sc-LDA+DMFT,LB08,IL15} to calculate the pressure-induced evolution of the electronic structure of paramagnetic MnO, FeO, CoO, and NiO oxides. 
It starts with construction of the effective low-energy (Kohn-Sham) Hamiltonian [$\hat{H}^{\mathrm{KS}}_{\sigma,\alpha\beta}(\bf{k})$] using the projection onto Wannier functions in order to obtain the $p$-$d$ Hubbard Hamiltonian (in the density-density approximation) \cite{MV97,Wannier-functions}
\begin{eqnarray}
\label{eq:hamilt}
\hat{H} = \sum_{\bf{k},\sigma} \hat{H}^{\mathrm{KS}}_{\sigma,\alpha\beta}({\bf{k}}) + \frac{1}{2} \sum_{\sigma\sigma',\alpha\beta} U_{\alpha\beta}^{\sigma\sigma'} \hat{n}_{\alpha\sigma} \hat{n}_{\beta\sigma'} - \hat{V}_{\mathrm{DC}},
\end{eqnarray}
where $\hat{n}_{\alpha\sigma}$ is the occupation number operator with spin $\sigma$ and (diagonal) orbital indices $\alpha$. $U_{\alpha\beta}^{\sigma\sigma'}$ denotes the reduced density-density form of the four-index Coulomb interaction matrix: $U_{\alpha\beta}^{\sigma\overline{\sigma}}=U_{\alpha\beta\alpha\beta}$ and $U_{\alpha\beta}^{\sigma\sigma}=U_{\alpha\beta\alpha\beta}-U_{\alpha\beta\beta\alpha}$. The latter is expressed in terms of the Slater integrals $F^0$, $F^2$, and $F^4$. For the $d$ electrons these parameters are related to the Coulomb and Hund's coupling as $U=F^0$, $J=(F^2+F^4)/14$, and $F^2/F^4=0.625$. $\hat{V}_\mathrm{DC}$ is the double-counting correction to account for the electronic interactions described by DFT (see below).

We use a fully self-consistent in charge density implementation of the DFT+DMFT method in order to take into account the effect of charge redistribution caused by electronic correlations and electron-lattice coupling \cite{charge-sc-LDA+DMFT}. To take into account self-consistency over charge density  we evaluate charge density $\rho( \mathbf{r} )$ as
\begin{eqnarray}
\label{eq:rho_scf}
\rho( \mathbf{r} )= k_BT \sum_{\mathbf{k},i\omega_n;ij} \rho_{ \mathbf{k};ij } G_{\mathbf{k};ji}( i\omega_n ) e^{i\omega_n0+},
\end{eqnarray}
where summation over the Matsubara frequencies is performed taking into account an analytically evaluated asymptotic correction.
In the plane-wave pseudopotential approach \cite{Pseudo} matrix elements of the charge density operator $\rho_{ \mathbf{k};ij }$ in the basis of the Kohn-Sham (KS) wave functions $\psi_{i\mathbf{k}}$ are defined as
\begin{eqnarray}
\rho_{\mathbf{k};ij}( \mathbf{r} )&=& \langle \psi_{i\mathbf{k}}| \mathbf{r} \rangle \langle \mathbf{r}|\psi_{j\mathbf{k}} \rangle \\ \nonumber &+& \sum_{Ilm} Q_{lm}^I({\bf r - R_I}) \langle \psi_{i\mathbf{k}}| {\beta^I_{l}} \rangle \langle {\beta^I_{m}} | \psi_{j\mathbf{k}} \rangle,
\end{eqnarray}
where $I$ is an atom index, $\beta^I_{m}( \mathbf{r} )$ is the augmentation basis function, $Q_{lm}^I( \mathbf{r} )$ is the augmentation function for charge density which is localized in the pseudopotential core ($|\mathbf{r}|< r_c$). Both $\beta^I_{l}( \mathbf{r} )$ and $Q_{lm}^I( \mathbf{r} )$ are calculated in an atomic calculation and are parameterized for a given pseudopotential. The lattice Green's function in the basis of KS wave functions is defined as 
\begin{eqnarray}
\label{eq:lattice_gf}
\hat{G}^{\sigma}(\mathbf{k},i\omega_n) = [(i\omega_n + \mu - \varepsilon^{\sigma}_{i\mathbf{k}}) \hat{I} -\hat{\Sigma}^{\sigma}(\mathbf{k},i\omega_n)]^{-1},
\end{eqnarray}
where $\varepsilon^{\sigma}_{i\mathbf{k}}$ are the KS eigenvalues calculated within DFT, $\hat{I}$ is the identity matrix. $\hat{\Sigma}^{\sigma}(\mathbf{k},i\omega_n)$ is the self-energy matrix computed from the solution of the effective impurity problem within DMFT by applying ``upfolding'' from the Wannier basis into the KS wave functions basis 
\begin{eqnarray}
\Sigma^{\sigma}_{ij}(\mathbf{k},i\omega_n) = \sum_{\nu \mu} P^{\sigma}_{i\nu}(\mathbf{k}) [\Sigma^{\sigma}_{\nu \mu}(i\omega_n) - V^\mathrm{DC}_{\nu\mu}] P^{\sigma*}_{i\mu}(\mathbf{k}),
\end{eqnarray}
where $P^{\sigma}_{i\nu}(\mathbf{k}) \equiv \langle \psi^{\sigma}_{i\mathbf{k}}|\hat{S}| \phi^\sigma _{\nu \mathbf{k}} \rangle$ are the matrix elements of orthonormal projection operator of the KS wave-functions $\psi^{\sigma}_{i\mathbf{k}}$ onto a basis set of atomic functions $\phi^\sigma _{\nu \mathbf{k}}$ with a given (in our case, $3d$) symmetry. $\hat{S}$ is an overlap operator in the ultrasoft pseudopotential scheme $\langle \psi_{ n\mathbf{k} }|\hat{S}|\psi_{n'\mathbf{k}}\rangle \equiv \delta_{nn'}$ given by
\begin{eqnarray}
S(\mathbf{r}, \mathbf{r'}) = \delta(\mathbf{r}-\mathbf{r'}) + \sum_{Ilm} q_{lm}^I\beta^I_{l}(\mathbf{r}-\mathbf{R}_I) \beta^{I*}_{m}(\mathbf{r'}-\mathbf{R}_I),
\end{eqnarray}
where $q^I_{lm} =\int Q_{lm}^I(\mathbf{r})d\mathbf{r}$ \citep{Pseudo}.
Note, for the non-correlated states matrix elements $P^{\sigma}_{i\nu}(\mathbf{k}) \equiv 0$ and hence $k_BT\sum_{i\omega_n}G_{\mathbf{k};ij} ( i\omega_n )e^{i\omega_n0+} = f_{i\mathbf{k}}\delta_{ij}$, where $f_{i\mathbf{k}}$ is the Fermi distribution function. In practice, it is useful to compute the total DFT+DMFT charge density  as  $\rho( {\bf r} ) = \rho_{\rm DFT}( \mathbf{r} ) + \Delta \rho( \mathbf{r} )$, i.e., to split the contribution from DFT and the charge-density correction due to electronic correlations in DMFT. Full charge self-consistency assumes both the convergence in the local self-energy
and in the electron density.
Since the Kohn-Sham energies from DFT already include interaction effects through the Hartree and exchange-correlation terms interaction contributions would be counted twice within DFT+DMFT. Therefore, to account for the electronic interactions already described by DFT we need to introduce a static correction to Hamiltonian in order to exclude the double-counting.
Here, we use the fully localized double-counting correction, evaluated from the self-consistently determined local occupations
$\hat{V}_{DC}=U ( N - \frac{1}{2} ) - J ( N_{\sigma} - \frac{1}{2} )$,
where $N_\sigma$ is the total $3d$ occupation with spin $\sigma$ and $N=N_\uparrow+N_\downarrow$.

To compute structural properties we evaluate total energy within DFT+DMFT as
\begin{eqnarray}
\label{eq:etot}
E= E_\mathrm{KS}[\rho(\mathbf{r})] + \langle \hat{H}^\mathrm{KS}\rangle - \sum_{i\mathbf{k}} \epsilon_{i\mathbf{k}}+ \langle \hat{H}_U\rangle - E_{DC}
\end{eqnarray}
where $E_\mathrm{KS}[\rho]$ is the KS total energy obtained for the self-consistent charge density $\rho(\mathbf{r})$ (Eq.~\ref{eq:rho_scf}). The 4-th term in Eq.~\ref{eq:etot} is the interaction energy $\langle H_U\rangle \equiv \frac{1}{2} \sum_{\sigma\sigma',\alpha\beta} U_{\alpha\beta}^{\sigma\sigma'} \langle \hat{n}_{\alpha\sigma} \hat{n}_{\beta\sigma'} \rangle$ computed from the double occupancy matrix evaluated within DMFT. $E_{DC}$ is the double-counting correction. In the case of the fully localized double-counting corrections, for a paramagnet it is evaluated as $E_{DC}=\frac{1}{2}UN(N-1)-\frac{1}{4}JN(N-2)$, where $N$ is a number of the Wannier $3d$ electrons. $\sum_{i\mathbf{k}} \epsilon_{i\mathbf{k}}$ is the sum of the KS valence-state eigenvalues which is evaluated as the thermal average of the KS low-energy Wannier Hamiltonian $\hat{H}^{\rm KS}_{\sigma, \mu \nu}(\mathbf{k}) = \sum_{ \varepsilon^\sigma_{i{\bf k}} \in [E_\mathrm{min},E_\mathrm{max}]} P^{\sigma*}_{i\mu}(\mathbf{k}) \varepsilon^{\sigma}_{i\mathbf{k}} P^{\sigma}_{i\nu}(\mathbf{k})$ with the non-interacting ($\hat{\Sigma}(i\omega_n) \equiv 0$) Green's function as
\begin{eqnarray}
\sum_{i\mathbf{k}} \epsilon_{i\mathbf{k}} = k_BT \sum_{\mathbf{k}\sigma\omega_n} \mathrm{Tr} [\hat{H}_\sigma^\mathrm{KS}(\mathbf{k}) \hat{G}^\sigma(\mathbf{k},i\omega_n)] e^{i\omega_n0^+}.
\end{eqnarray}
Here, $E_\mathrm{min}$ and $E_\mathrm{max}$ define the energy range window to compute the Wannier functions that are treated as correlated orbitals.
$\langle \hat{H}^\mathrm{KS}\rangle$ is evaluated similarly but with the full Green's function including
the self-energy. To calculate these two contributions, the summation is performed over the Matsubara frequencies with analytically evaluated asymptotic correction \cite{IL15}. Using this approach, we can determine correlation-induced structural transformations, as well as the corresponding change in the atomic coordinates and of the unit-cell shape. 

\section*{III. COMPUTATIONAL DETAILS}

We employ the DFT+DMFT approach \cite{charge-sc-LDA+DMFT,LB08,IL15} as discussed above to explore the electronic structure and magnetic properties of MnO, FeO, CoO, and NiO under pressure.
For the partially filled Mn, Fe, Co, and Ni $3d$ and O $2p$ orbitals we construct a basis set of atomic-centered symmetry-constrained Wannier functions, defined over the energy range window spanned by the $p$-$d$ band complex \cite{MV97,Wannier-functions}.
We employ the continuous-time hybridization-expansion quantum Monte-Carlo algorithm \cite{ctqmc} in order to solve the realistic many-body problem.
The calculations are performed in the paramagnetic state at an electronic temperature $T = 1160$ K, i.e., well above the N\'eel temperature, ranging from $T_N\sim 116$ K to 523 K for MnO to NiO. At such a temperature MnO-NiO oxides adopt a cubic B1 crystal structure up to high pressures.
We take the following values of the average Hubbard $U$ and Hund's exchange $J$ as estimated  previously: $U=8.0$ eV and $J=0.86$ eV for the Mn $3d$ orbitals, 7.0 and 0.89 eV for Fe, 8.0 and 0.9 eV for Co, and 10.0 and 1.0 eV for Ni, respectively \cite{band_structure,KL08,IL15,BK12,TM10}. The Coulomb interaction $U$ and Hund's $J$ are considered to be pressure-independent and have been treated in the density-density approximation. The spin-orbit coupling
was neglected in these calculations.
In out DFT+DMFT calculation we neglect by the effect of lattice and local magnetic moments entropy on the equation of states of MnO-NiO which seems to become prominent at very high temperatures $T \gg 1160$ K \cite{Greenberg}.
The spectral functions were computed using the maximum entropy method and the Pad\'e analytical continuation procedure.


\section*{IV. RESULTS AND DISCUSSION}

\begin{figure}[tbp!]
\centerline{\includegraphics[width=0.5\textwidth,clip=true]{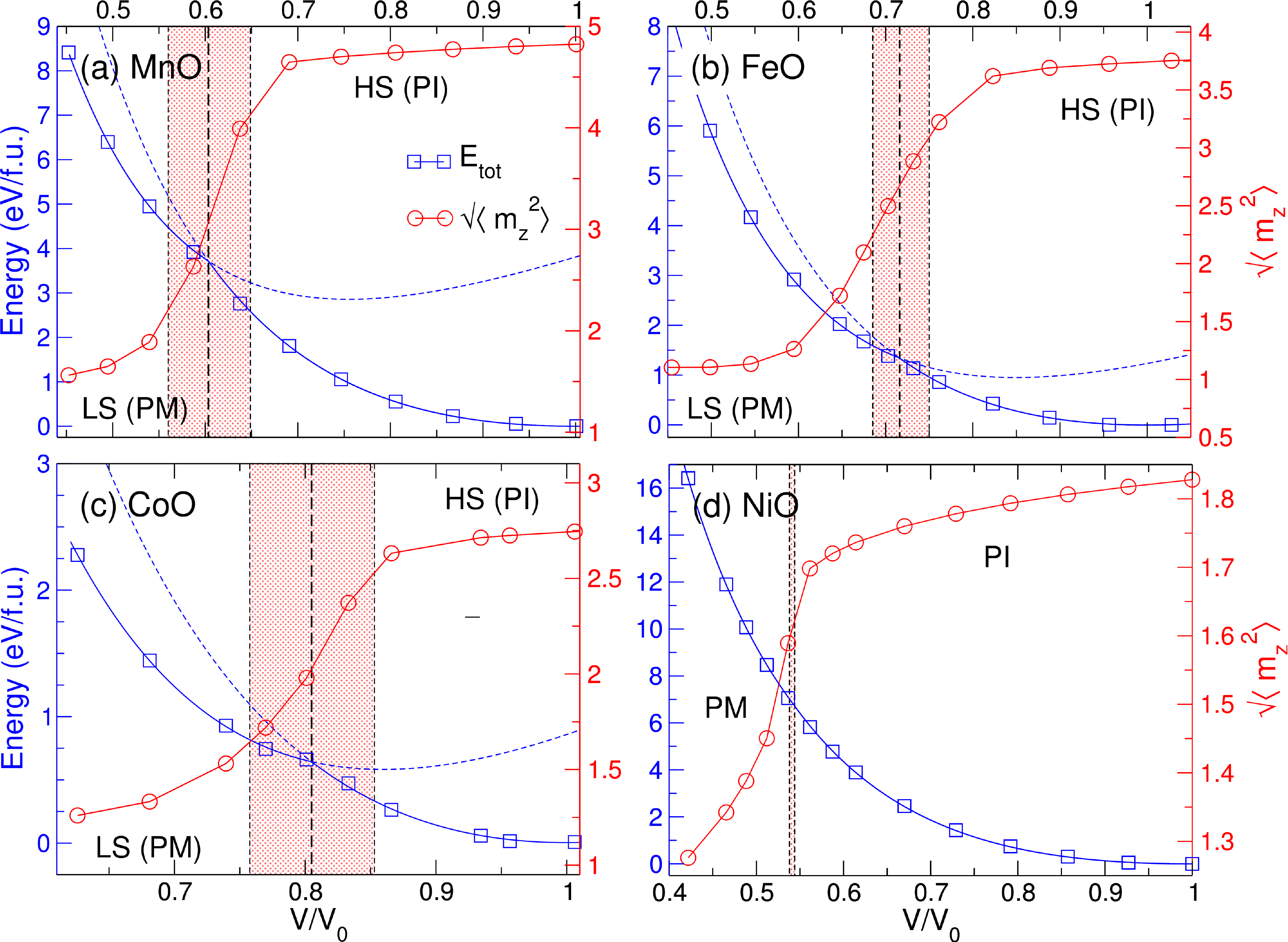}}
\caption{Evolution of the total energy and local magnetic moments $\sqrt{ \langle \hat{m}_z^2 \rangle}$ of paramagnetic MnO, FeO, CoO, and NiO obtained by DFT+DMFT as a function of relative volume $V/V_0$. $V_0$ is the calculated equilibrium lattice volume. The calculated equilibrium lattice volume $V \simeq 158.9$, 144.1, 137, and 128 a.u.$^3$ for MnO, FeO, CoO, and NiO, respectively. Our result for the lattice volume collapse evaluated from the Maxwell construction for the DFT+DMFT total energy results is marked by a red shaded rectangle. 
}
\label{Fig_1}
\end{figure}

As a starting point, we calculate the total energy and fluctuating (instantaneous) local moments $\sqrt{\langle \hat{m}^2_z \rangle}$ of the paramagnetic B1-structured MnO, FeO, CoO, and NiO as a function of the unit-cell volume using the DFT+DMFT method (see Figure~\ref{Fig_1}) \cite{IL15}. Overall, the calculated electronic, magnetic, and lattice properties of MnO-NiO agree well with those published previously \cite{IL15,BK12,KL08}. We obtain  $V \simeq 158.9$, 144.1, 137, and 128 a.u.$^3$ for the equilibrium lattice volume of MnO, FeO, CoO, and NiO, respectively. The ambient-pressure local magnetic moments of $\sim$1.8$\mu_B$–-4.8$\mu_B$ in NiO-MnO match the high-spin magnetic state of their transition metal ions. The local moments are seen to retain the HS state upon compression down to
about 0.6-0.7 $V/V_0$. Upon further compression, all these compounds exhibit magnetic collapse -- a remarkable reduction of the local magnetic moments which results in a Mott insulator-to-metal phase transition (IMT) \cite{IL15}. The resulting low-spin local moments are about 1.6$\mu_B$, 1.1$\mu_B$, 1.26$\mu_B$, and 1.28$\mu_B$ for MnO, FeO, CoO, and NiO, respectively. By fitting the DFT+DMFT total-energy results to the third-order Birch-Murnaghan equation of states (separately for the low- and high-volume regions) we obtain that magnetic collapse is accompanied by a sudden change of the lattice volume. That is, the phase transition is of first order with a significant fractional volume $\Delta V/V$ collapse of 13.6\%, 9\%, and 11.2 \% for MnO-CoO, except for NiO, where a resulting change of the lattice volume is only about 1.4\%. 

Our result for the transition pressure $p_c$ evaluated from the equation of states shows a monotonous decrease from 145 GPa, 73 GPa, to 40 GPa for MnO, FeO and CoO, respectively, while NiO has a high transition pressure $\sim$429 GPa.
We note that this anomalous behavior of $p_c$ can be understood as a
continuous decrease of the strength of electronic correlations and, hence, the tendency towards localization of
the $3d$ electrons upon changing of the electron configuration from $3d^5$ 
Mn$^{2+}$ ions to $3d^7$ in Co$^{2+}$ \cite{IL15}. In fact, the effective interaction strength changes from $U + 4J$ for MnO to $U - 3J$ for CoO, while in NiO it sharply increases
due to a crossover in the effective degeneracy of low-energy
excitations from five-orbital (as in MnO, FeO, and CoO) to
two-orbital behavior (as in NiO). 
It is interesting to note that the calculated transition pressure $p_c$ (as well as the energy gap values) is very sensitive to the choice of the interaction
parameters Hubbard $U$ and Hund's exchange $J$. To help check this result, we perform the DFT+DMFT calculations of MnO-NiO with various sets of the Hubbard $U$ and Hund's $J$ values, varying them by about 10\%. In particular, for MnO, we obtain a transition pressure $p_c \sim 145$ GPa, 133 GPa, and 109 GPa for the $U/J$: 8/0.86 eV, 7/0.86 eV, and 8/
0.75 eV, respectively. For FeO, it is $p_c \sim 55$ GPa, 73 GPa, and 80 GPa for the $U/J$: 5/0.89 eV, 7/0.89 eV, and 8/0.89 eV, respectively. For NiO our results are $\sim$248 GPa and 429 GPa respectively for the $U/J$: 8/1 eV and 10/1 eV. In fact, our results are consistent with the behavior the effective interaction strength that changes from $U + 4J$ for MnO to $U - 3J$ for CoO, revealing a strong sensitivity (in terms of a percentage change) to $J$ rather than to $U$. 

Overall, our results suggest a complex interplay between chemical bonding and electronic correlations in MnO-NiO near the Mott transition. The Mott insulator-to-metal phase transition is accompanied by a remarkable increase of the bulk modulus, varying from 137 to 263 GPa in MnO, 142/210 GPa in FeO, 184/246 GPa in CoO, and 187/188 GPa in NiO. The latter implies an anomaly in the compressibility at the phase transition point. Moreover, we obtain a substantial redistribution of electrons between the $t_{2g}$ and $e_g$ orbitals in the $3d$ shell of the spin-state active MnO, FeO, and CoO. This implies a significant change in chemical bonding of the $3d$ electrons. Thus, the $t_{2g}$ orbital occupations are found to gradually increase with pressure, whereas the $e_g$ orbitals are strongly depopulated (below 0.27 for MnO and FeO, and 0.44 for CoO). The $3d$ total occupancy weakly changes with pressure. Upon pressurizing to $\sim$150 GPa it increases by 0.4-0.5 electrons in MnO, FeO, and CoO.
It is important to note that
magnetic collapse is also seen to occur in NiO, which in fact, has a $3d^8$ electronic
configuration of the Ni$^{2+}$ ion with completely occupied $t_{2g}$ and half-filled $e_g$
bands. 

\begin{figure}[tbp!]
\centerline{\includegraphics[width=0.5\textwidth,clip=true]{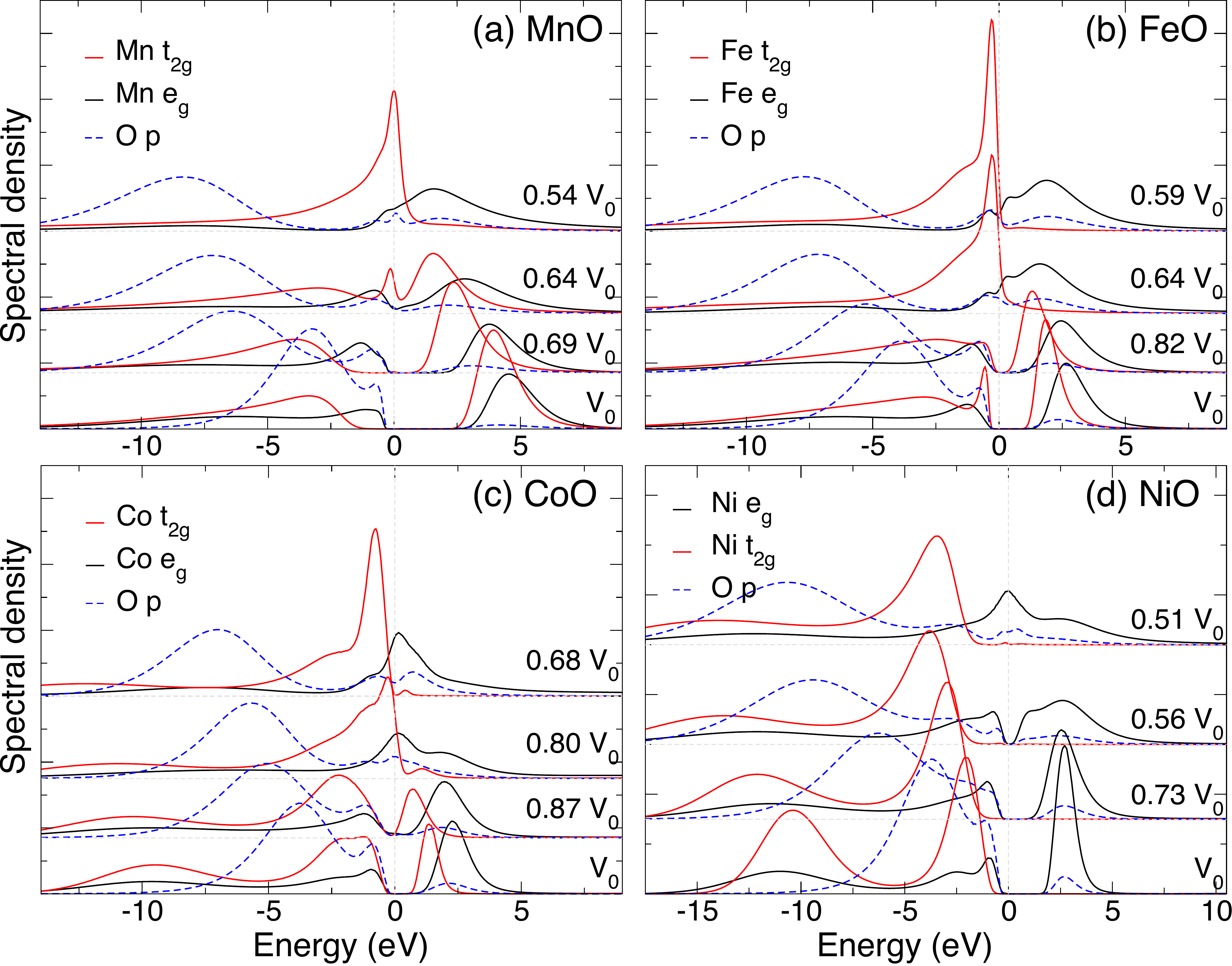}}
\caption{Evolution of the transition metal $t_{2g}$ and $e_g$ and oxygen $2p$ spectral functions of paramagnetic MnO, FeO, CoO, and NiO calculated by DFT+DMFT as a function of relative volume $V/V_0$. }
\label{Fig_2}
\end{figure}

\begin{figure}[tbp!]
\centerline{\includegraphics[width=0.5\textwidth,clip=true]{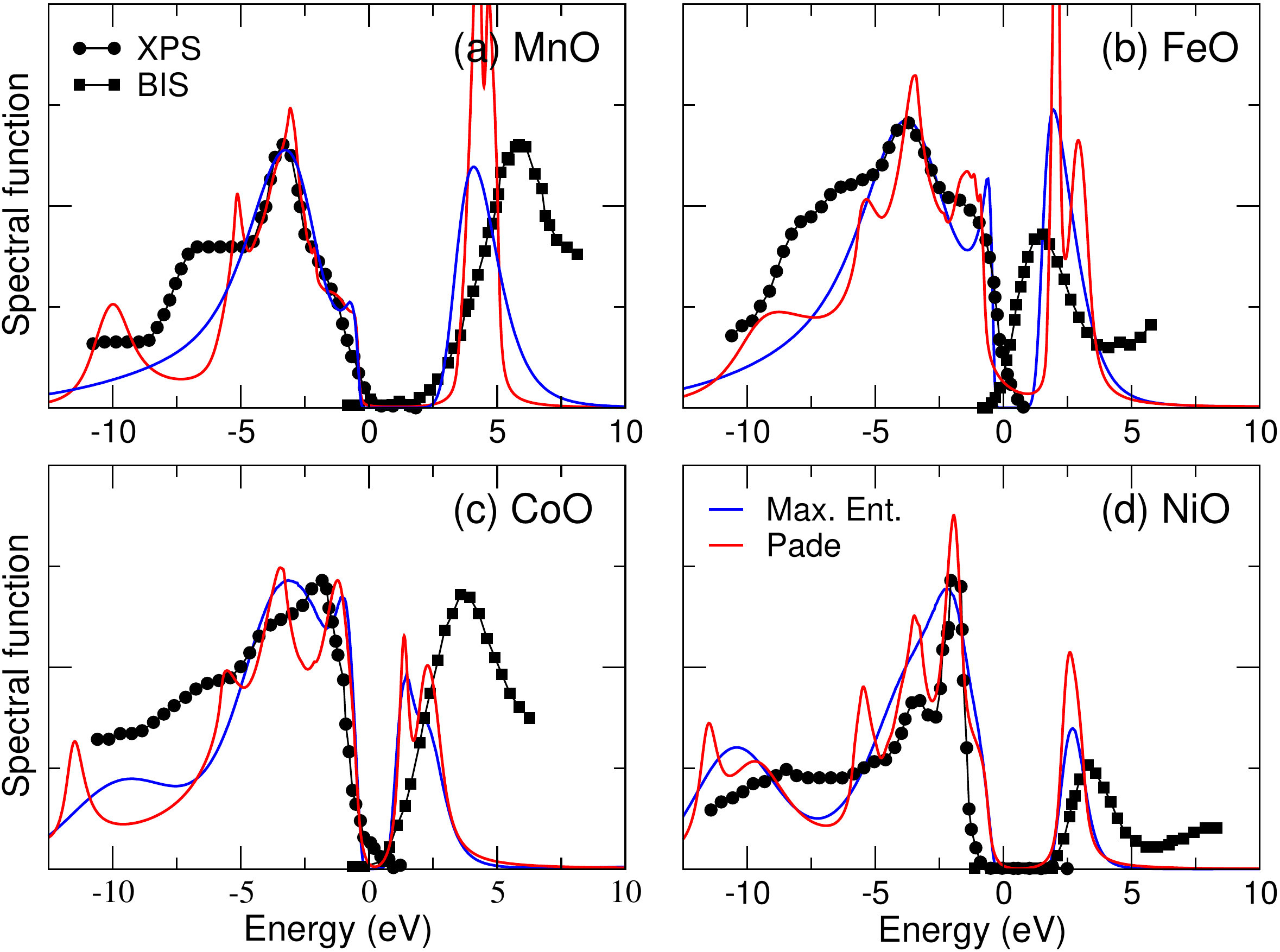}}
\caption{Comparison of theoretical spectral functions obtained by DFT+DMFT to experimental XPS and BIS spectra. XPS and BIS results are extracted from Ref.~\cite{exp_spectra}. Max.~Ent. stands for the spectral functions
computed using the maximum entropy method. Pade -- for the spectral functions
evaluated using the Pad\'e analytical continuation procedure.}
\label{Fig_3}
\end{figure}

\begin{figure}[tbp!]
\centerline{\includegraphics[width=0.45\textwidth,clip=true]{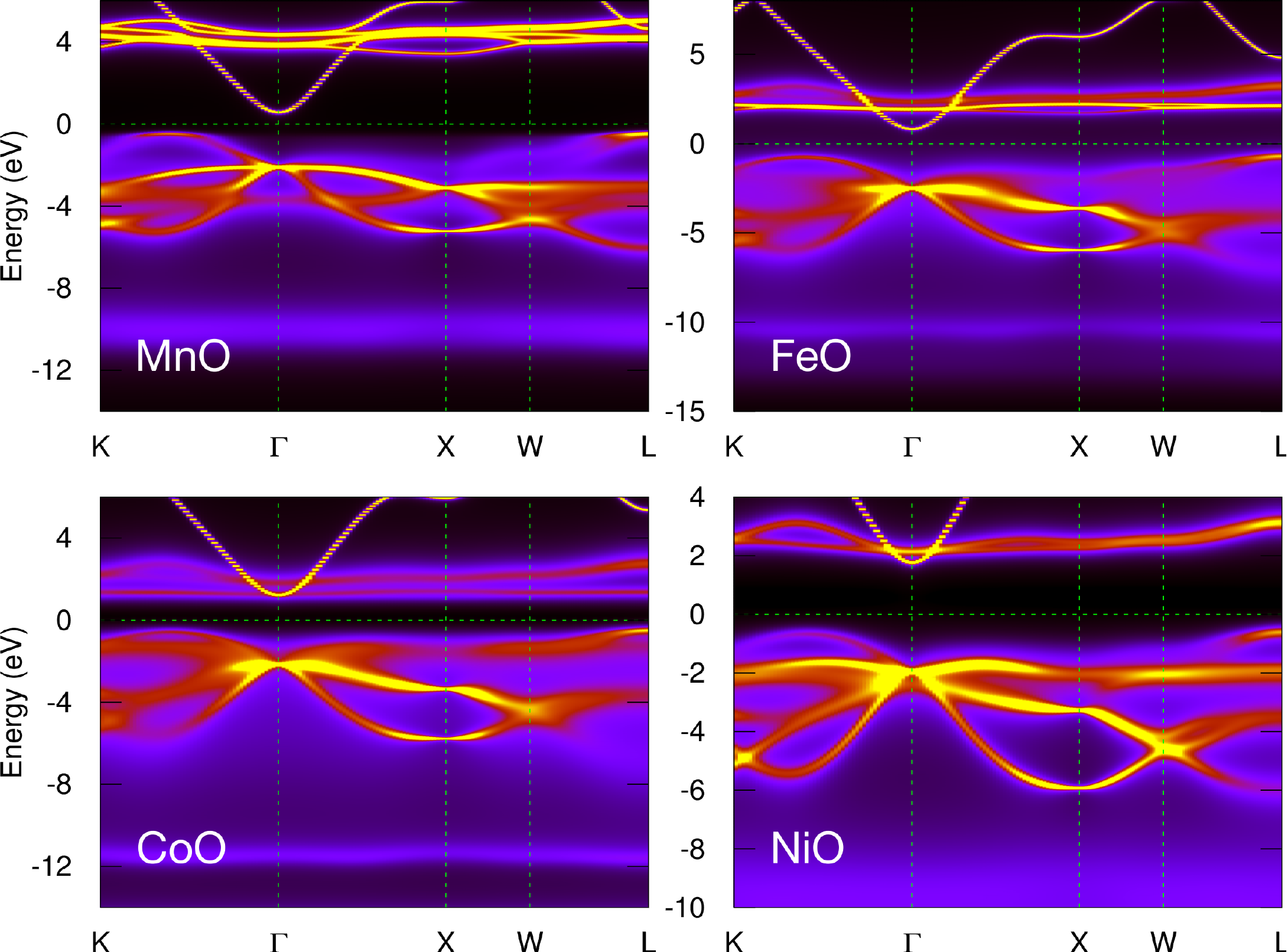}}
\caption{{\bf k}-resolved spectral function of paramagnetic
MnO, FeO, CoO, and NiO computed within DFT+DMFT for the Mott insulating phase at the equilibrium volume $V_0$. 
}
\label{Fig_4}
\end{figure}

\begin{figure}[tbp!]
\centerline{\includegraphics[width=0.45\textwidth,clip=true]{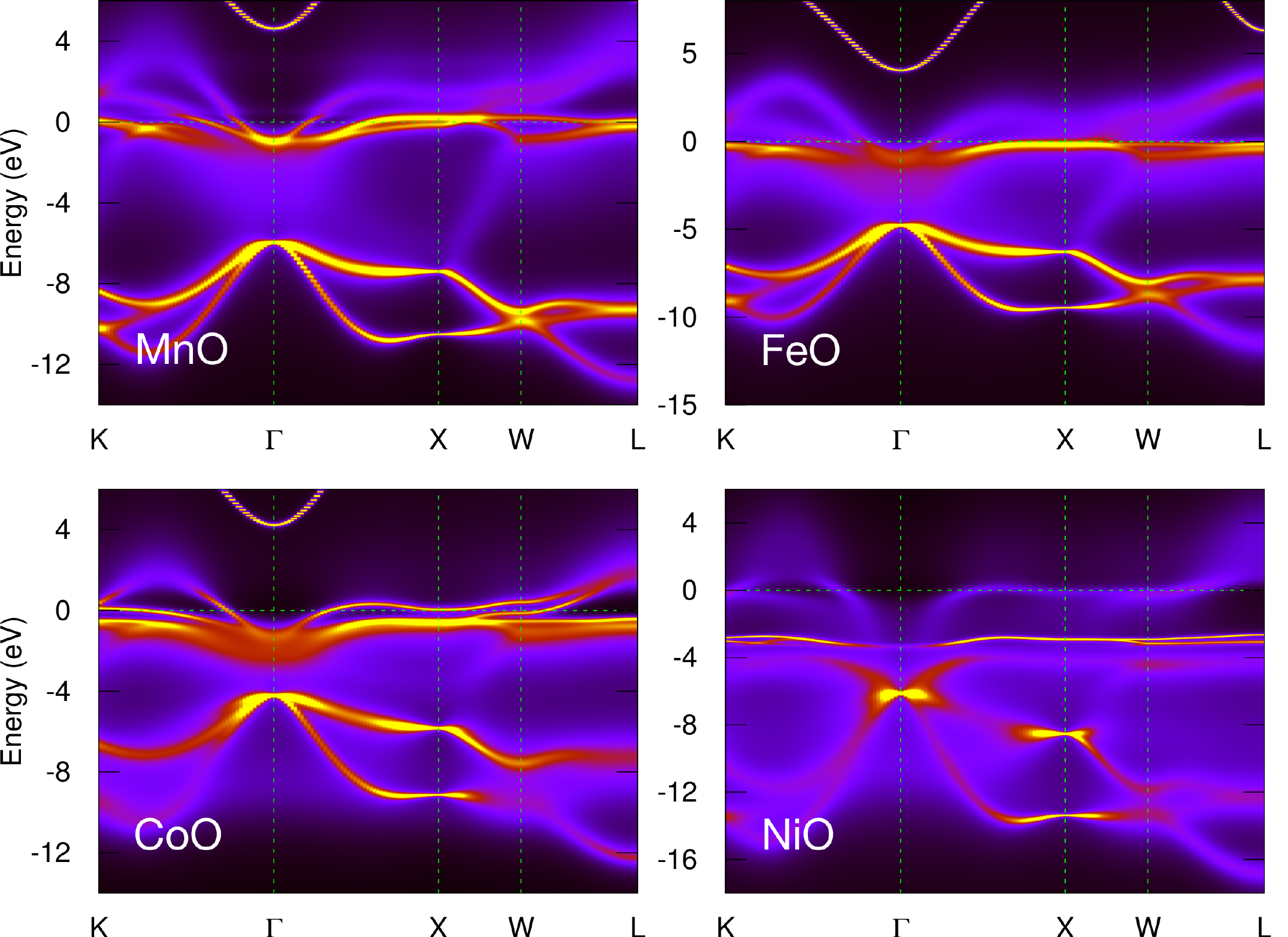}}
\caption{{\bf k}-resolved spectral function of paramagnetic
MnO, FeO, CoO, and NiO computed by DFT+DMFT for the metallic phase with the lattice constant $a=7.0$, 7.2, 7.2, and 6.4 a.u. (for the lattice volume $V/V_0=0.54$, 0.64, 0.68, and 0.51) for MnO, FeO, CoO, and NiO, respectively.
}
\label{Fig_5}
\end{figure}

In Figure~\ref{Fig_2} we present our results for the evolution of the transition metal $t_{2g}$ and $e_g$ and oxygen $2p$ spectral functions of MnO, FeO, CoO, and NiO obtained by DFT+DMFT as a function of compression. 
A comparison of our results to experimental XPS and BIS spectra are shown in Fig.~\ref{Fig_3}. 
In agreement with experimental data, at ambient pressure all these compounds (from MnO to NiO)
are Mott insulators with a large $d$-$d$ energy gap of $\sim$2--3.5 eV, which emphasizes the crucial importance of strong correlations to determine the electronic and magnetic properties of transition metal oxides.
The top of the valence band is predominantly formed by transition metal $3d$ states, with a large contribution from the O $2p$ states. 
In Figures~\ref{Fig_4} and \ref{Fig_5}  we display our results for the {\bf k}-resolved spectral functions of paramagnetic MnO-NiO obtained by DFT+DMFT for the ambient pressure Mott insulating phase and those for pressure $p>p_c$ in a metallic state. It is important to note the contribution of the empty transition metal $4s$ states at the Brillouin zone $\Gamma$-point seen as a broad parabolic-like band above the Fermi energy (see Figure~\ref{Fig_4}). This is in agreement
with photoemission and optical experiments which e.g. for FeO report a weak absorption between 0.5 and 2.0 eV, assigned to the mixed Fe $3d$-O $2p$ to Fe $4s$ transitions, while the strong absorption edge associated with the $d$-$d$ transitions is found to appear in optical spectroscopy at about 2.4 eV \cite{optics}.

\begin{figure}[tbp!]
\centerline{\includegraphics[width=0.45\textwidth,clip=true]{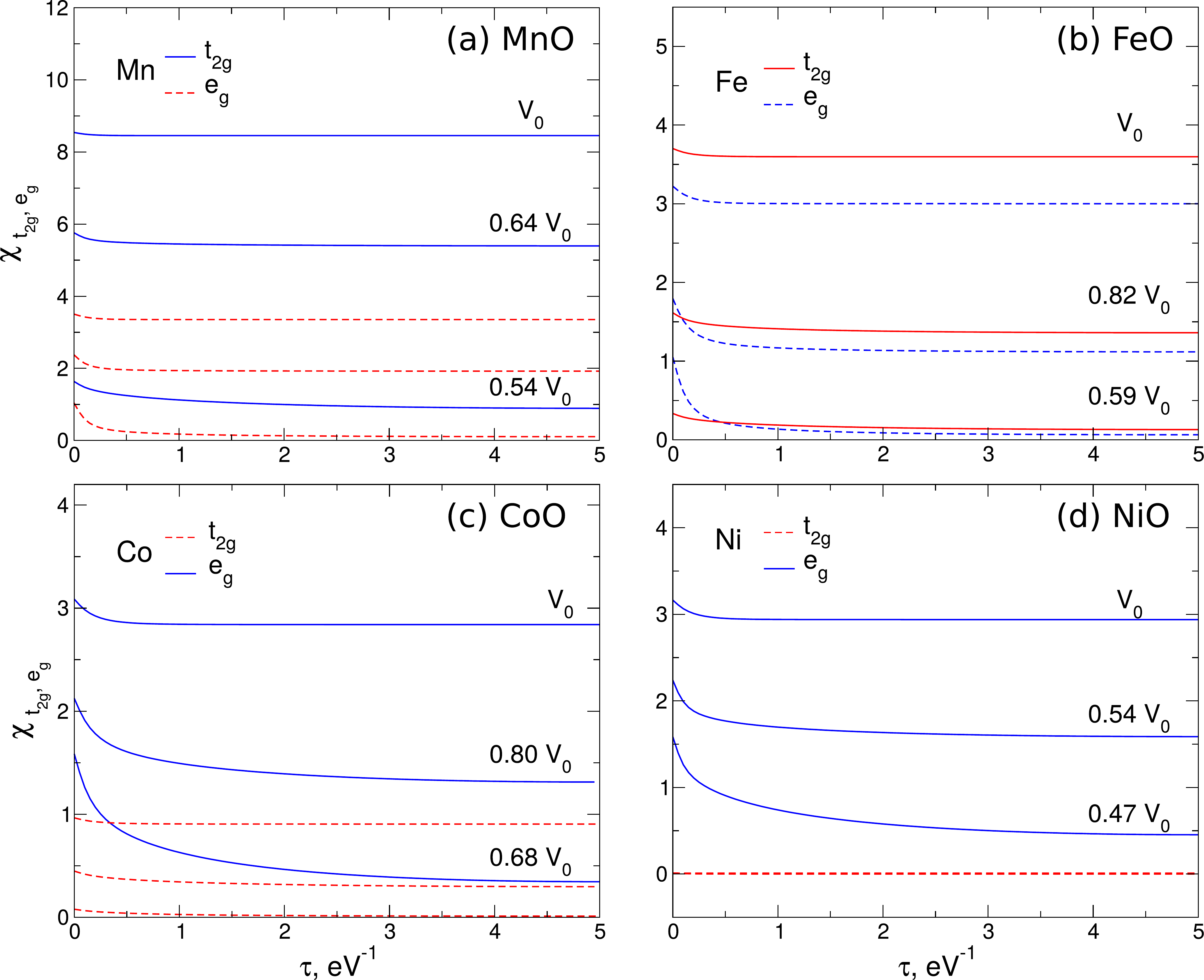}}
\caption{Local spin-spin correlation function $\chi(\tau)=\langle\hat{m}_z(\tau)\hat{m}_z(0)\rangle$ of paramagnetic MnO, FeO, CoO, and NiO calculated by DFT+DMFT for different volumes. $\tau$ is the imaginary time. The intraorbital $t_{2g}$ and $e_g$ contributions are shown.
}
\label{Fig_6}
\end{figure}

Under pressure the energy gap in MnO-NiO gradually decreases resulting in a Mott insulator-to-metal phase transition. 
Upon the Mott transition, all these materials exhibit a strongly correlated metallic behavior. It is characterized by the lower- and upper-Hubbard bands to appear in their spectral function, and
the quasiparticle peak at the Fermi level, associated with a
substantial renormalization of the electron mass (see Figure~\ref{Fig_2}).
We find that the electronic effective mass evaluated by using a polynomial fit of the imaginary part of the self-energy $\Sigma(i\omega_n)$ at the lowest Matsubara frequencies $\omega_n$ diverges at the Mott transition (upon decompression starting from the metallic phase), in accordance with a Brinkman-Rice picture of the Mott IMT \cite{PhysRevB.2.4302}. We note that this divergence concurs with a drop of the spectral weight of the $t_{2g}$ and $e_g$ orbitals at the Fermi level (with an opening of a Mott energy gap) and a sudden increase of the local magnetic moments in MnO-NiO.
Our analysis of the spectral weight at the Fermi level and the quasiparticle weights suggests that the Mott IMT is accompanied by a \emph{simultaneous} collapse of magnetic moments and lattice volume. 
The latter clearly indicates the crucial importance of
electronic correlations of localized $3d$ electrons to explain the electronic structure and
lattice properties of correlated transition metal oxides.
Indeed, at ambient pressure, the $3d$ electrons are strongly localized, as
it is seen from our result for the local spin susceptibility
$\chi(\tau)=\langle\hat{m}_z(\tau)\hat{m}_z(0)\rangle$, where $\tau$ is the imaginary time (see
Figure~\ref{Fig_6}). Indeed, $\chi(\tau)$ is seen to be nearly constant, independent on $\tau$. Upon further compression, the $3d$ electrons exhibit
a crossover from localized to itinerant moment behavior which is associated with a Mott transition, as it is clearly
seen in paramagnetic FeO, CoO, and NiO. In particular, $\chi(\tau)$ is seen to decay fast nearly to zero with the imaginary time $\tau$, which is typical for itinerant magnets.

\begin{figure}[tbp!]
\centerline{\includegraphics[width=0.45\textwidth,clip=true]{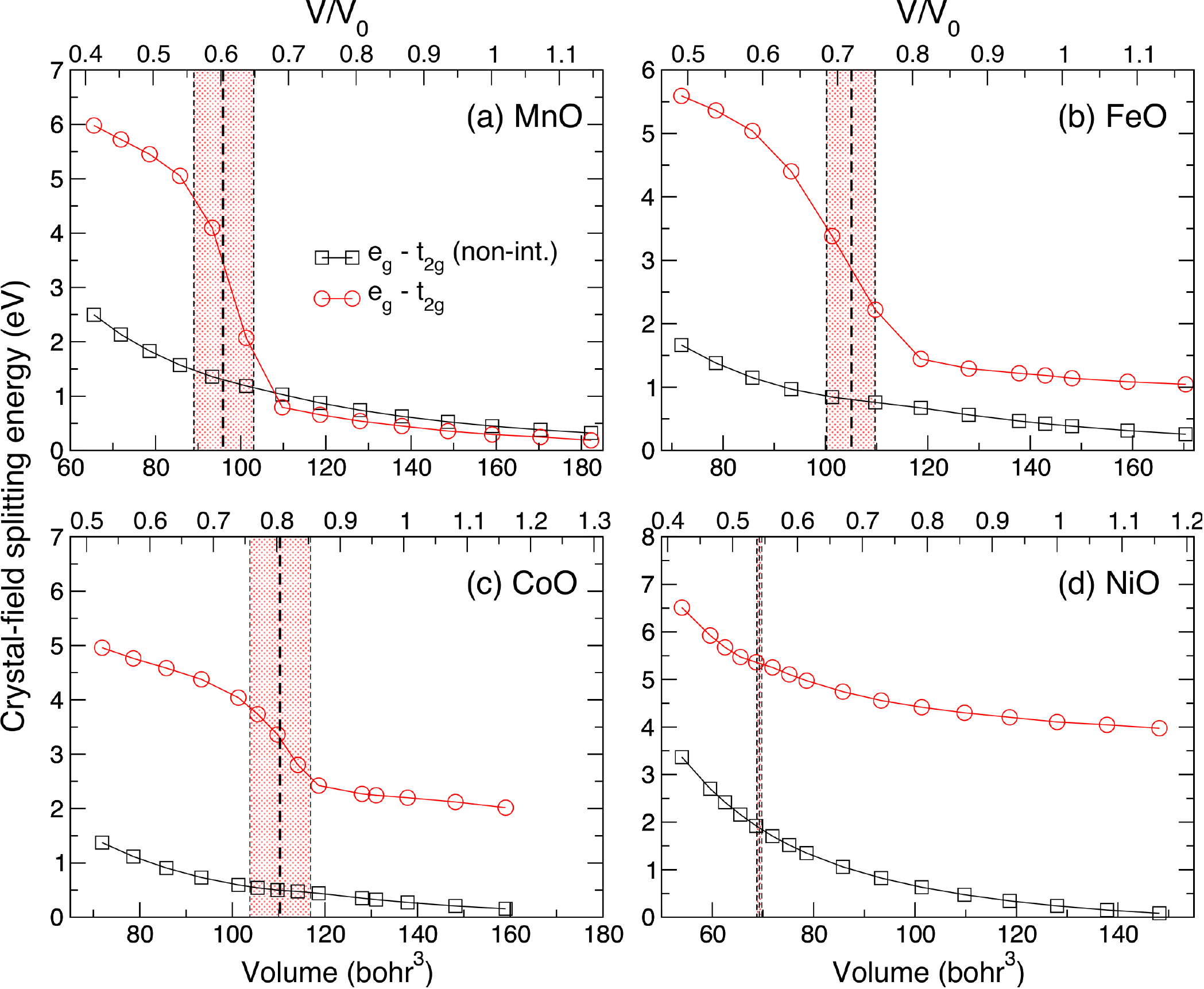}}
\caption{Evolution of the effective $t_{2g}$-$e_g$ crystal field splittings of the Wannier $3d$ orbitals obtained by DFT+DMFT for paramagnetic MnO, FeO, CoO, and NiO as a function of lattice volume. 
}
\label{Fig_7}
\end{figure}

\begin{figure}[tbp!]
\centerline{\includegraphics[width=0.45\textwidth,clip=true]{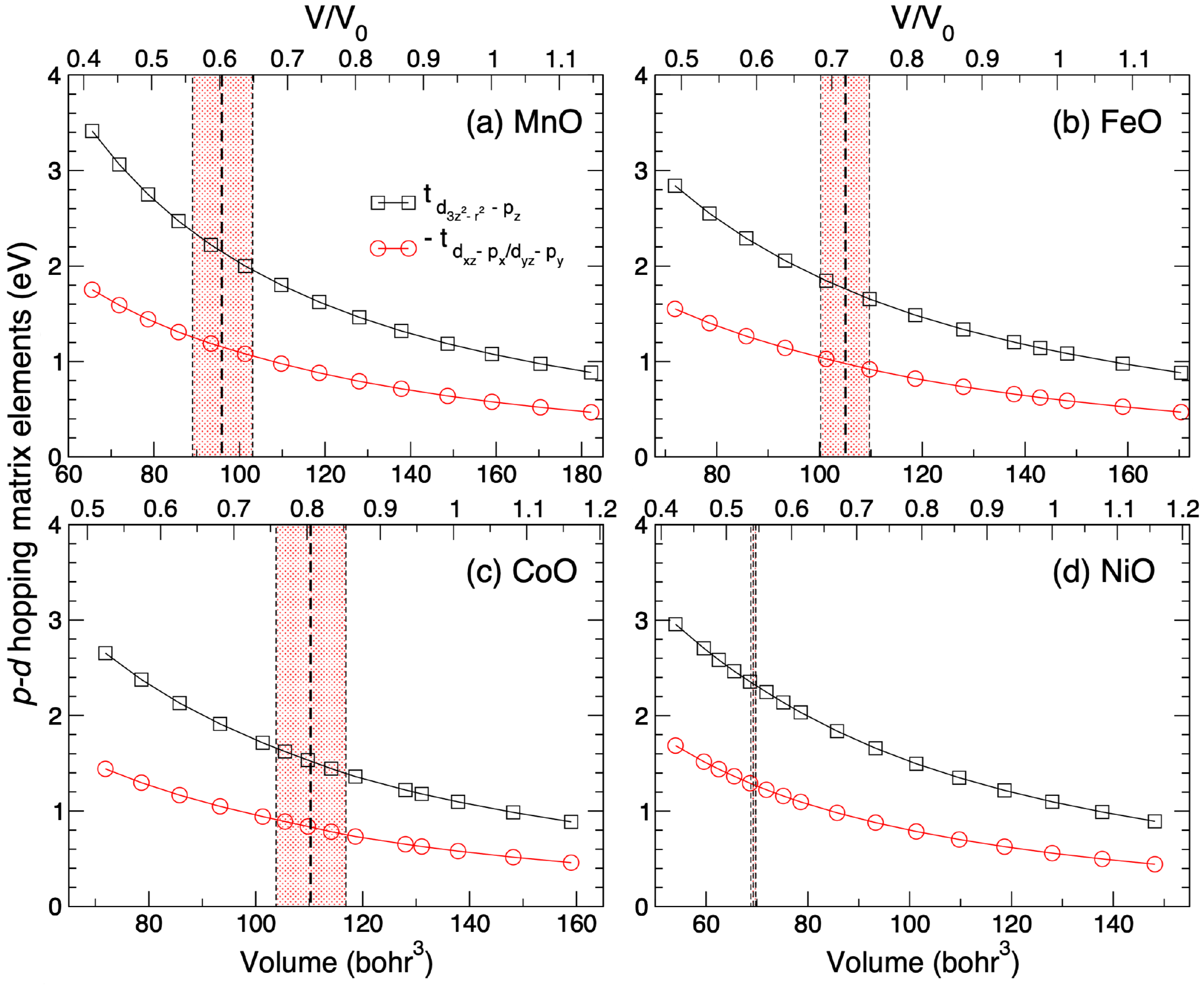}}
\caption{Orbitally-resolved $p$-$d$ hopping matrix elements of paramagnetic MnO, FeO, CoO, and NiO calculated by DFT+DMFT for different lattice volume. 
}
\label{Fig_8}
\end{figure}

In Figures~\ref{Fig_7}  and \ref{Fig_8} we display our results for the $t_{2g}$-$e_g$ crystal field energy splitting and the $p$-$d$ hopping matrix elements of MnO-NiO as a function of volume. The crystal field splittings are obtained from the first moments of the interacting lattice Green's function for the $3d$ states as $\Delta_{t_{2g}\_e_g} \equiv \mathrm{diag}[\sum_{\bf k}H^\mathrm{KS}({\bf k})+\mathrm{Re}\Sigma(i\omega_n \rightarrow \infty) ]$, where $H^\mathrm{KS}({\bf k})$ is the effective low-energy $p$-$d$ (Kohn-Sham) Hamiltonian in the Wannier basis set.
$\mathrm{Re}\Sigma(i\omega_n \rightarrow \infty)$ is a static Hartree contribution from self-energy $\Sigma(i\omega_n)$. We also compare our results for $\Delta_{t_{2g}\_e_g}$ with those obtained in the non-interacting case, with
$\Sigma(i\omega_n) \equiv 0 $. We observe that upon compression both the non-interacting $t_{2g}$-$e_g$ crystal field energy splittings and the $p$-$d$ ($p_z$-$d_{3z^2-r^2}$ and $p_x$-$d_{xz}$/$p_y$-$d_{yz}$) hopping matrix elements monotonously increase (by modulus) (i.e., as expected the transition-metal $3d$ bandwidth and $t_{2g}\_e_g$ splitting monotonously increase under pressure). We note that neither $\Delta_{t_{2g}\_e_g}$ for $\Sigma(i \omega_n) \equiv 0$ nor $p$-$d$ hopping parameters exhibit anomaly (are changing continuously) near the Mott transition. In contrast to that the Mott IMT accompanied by the HS-LS transition clearly correlates with a remarkable enhancement of the crystal-field splitting, caused by correlation effects. This change of $\Delta_{t_{2g}\_e_g} $ is large, about 1.5-3.2 eV for MnO-CoO, whereas for NiO it is seen as a weak anomaly at the transition point. This result implies the crucial importance of electronic correlation effects, determined by the self-energy contribution $\Sigma(i\omega_n)$, which plays a significant role at the Mott IMT. 
Our results are consistent with a transition from localized to itinerant
moment behavior of the $3d$ electrons at the Mott transition, in which the Mott IMT
concurs with a collapse of magnetism. Indeed, under pressure, the overlap of the transition metal $3d$ and ligand $2p$ orbitals increases and hence the $p$-$d$ hybridization (and $3d$ bandwidth) increases, resulting in a reduction of correlation effects and metallization for $p>p_c$.
This behavior concurs with an increase of crystal field splitting between the $t_{2g}$ and $e_g$ orbitals which favors the lower spin state. 

Next, we analyze the pressure-evolution of the electronic structure of MnO-NiO in the vicinity of the Mott IMT. At this point, we determine a reduced density matrix of the $3d$ electrons as a function of lattice volume: $\rho_{\alpha\beta}=|\phi^{N,S_z}_\alpha \rangle w_{N,S_z} \langle \phi^{N,S_z}_\beta|$, where $\phi^{N,S_z}_\alpha$ is a $3d$ atomic state with the occupation $N$ and spin $S_z$ \cite{fluctuations}. Its eigenvalues $w_{N,S_z}$ give a probability of observing different $3d$-electron atomic configurations for a given unit-cell volume. That is, the $3d$ electrons are seen being fluctuating between various atomic configurations with a given probability, exchanging with the surrounding medium, that gives alternative information about the nominal valence ($\mathrm{Tr}_{ \{ S_z \} }w_{N,S_z}$) and spin-state ($\mathrm{Tr}_{ \{ N \} }w_{N,S_z}$). 

\begin{figure}[tbp!]
\centerline{\includegraphics[width=0.5\textwidth,clip=true]{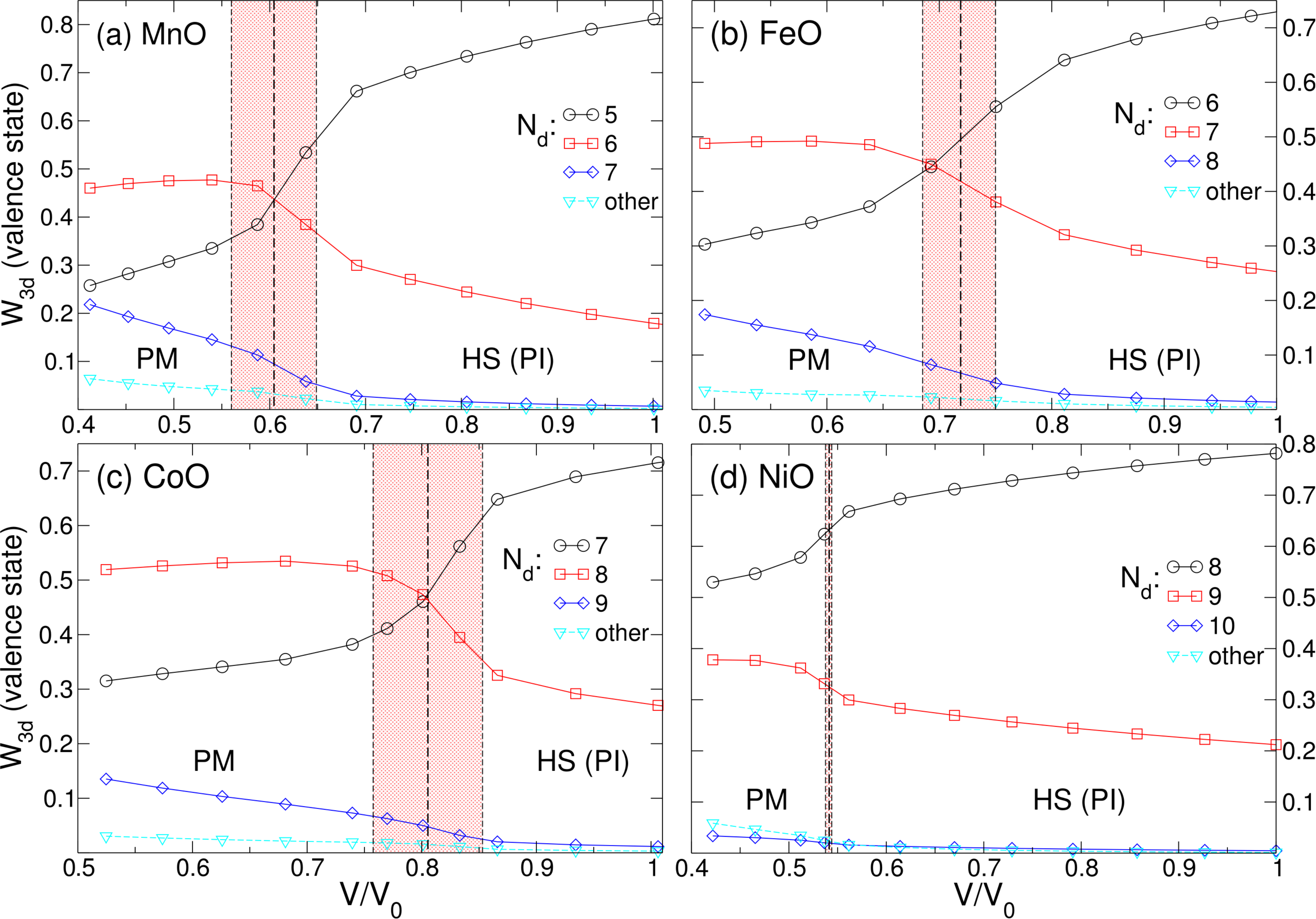}}
\caption{Evolution of the calculated valence states of MnO, FeO, CoO, and NiO calculated by DFT+DMFT as a function of relative volume $V/V_0$. $N_d$ denotes the corresponding $3d^{N_d}$ electronic configurations.
}
\label{Fig_9}
\end{figure}

\begin{figure}[tbp!]
\centerline{\includegraphics[width=0.5\textwidth,clip=true]{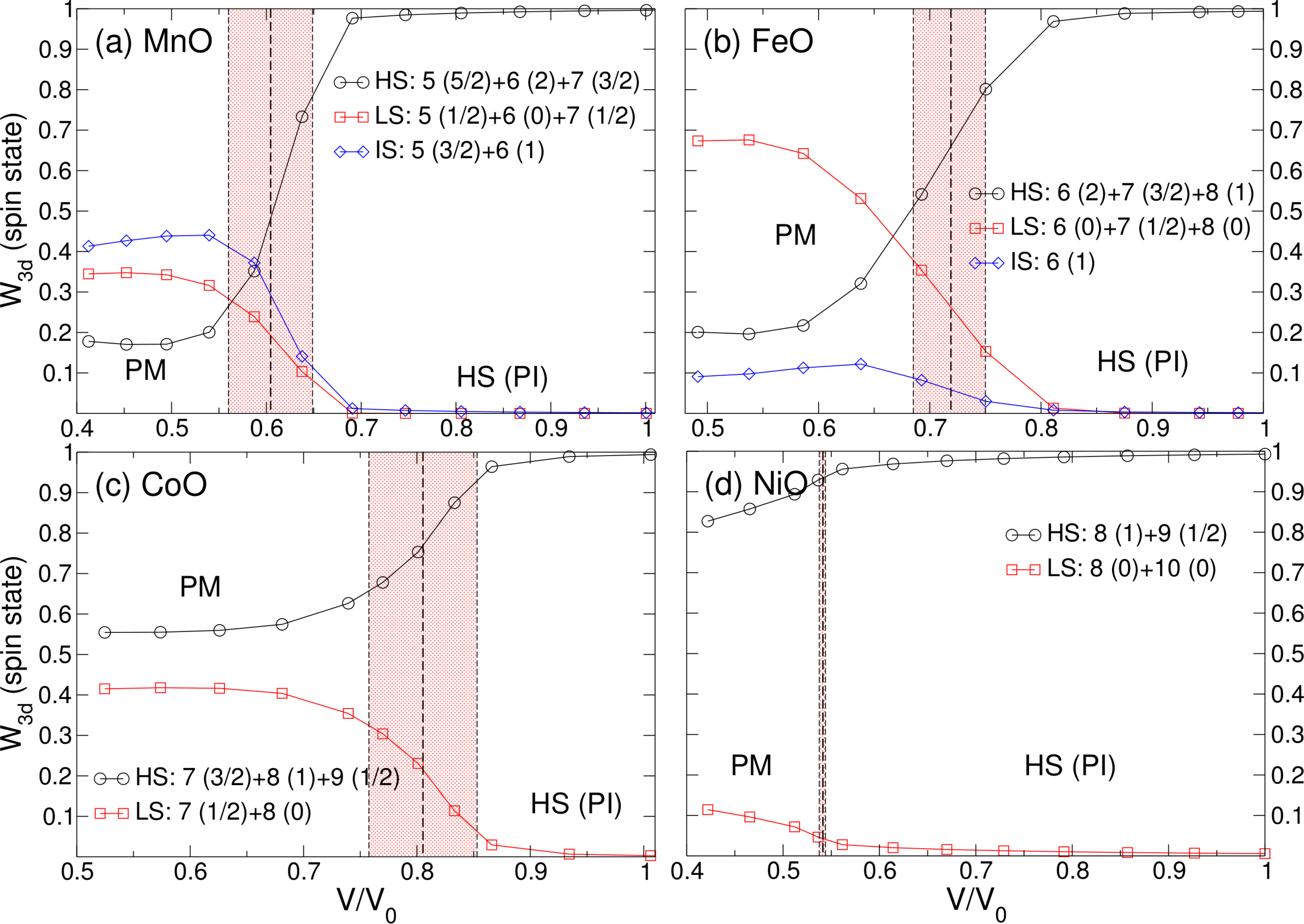}}
\caption{
Our DFT+DMFT results for the spin-state configurations weights of MnO, FeO, CoO and NiO as a function of relative volume $V/V_0$. Different spin-state contributions, e.g., the HS: $|d^5_{S_z=5/2}\rangle+|d^6_{S_z=2} \rangle+|d^7_{S_z=3/2} \rangle$, IS: $|d^5_{S_z=3/2}\rangle+|d^6_{S_z=1}\rangle$, and LS: $|d^5_{S_z=1/2} \rangle+|d^6_{S_z=0}\rangle+|d^7_{S_z=1/2} \rangle$ for MnO, are shown.
}
\label{Fig_10}
\end{figure}

Our results for the pressure-evolution of the calculated valence state and spin-state configurations weights of MnO-NiO are summarized in Figures~\ref{Fig_9} and \ref{Fig_10}. Our DFT+DMFT calculations reveal that upon moderate compression (in the Mott insulating regime) all four compounds adopt a 2+ oxidation state of their transition-metal ions with a nominal $3d^N$ configuration with $N$ varying from 5 to 8 for MnO-to-NiO, respectively. In fact, at ambient pressure, the nominal 2+ oxidation state has the largest weight of above 70-80\%. It reduces to about $\leq 60$\% upon compression to 0.6-0.7~$V/V_0$, all in the Mott insulating regime. Interestingly, at ambient pressure the weight of the first excited state corresponding to the $3d^{N+1}$ configuration is below 20-30\%. However, it tends to increase upon compression, resulting in a remarkable crossover of the valence state of MnO-CoO. 
In fact, as shown in Figure~\ref{Fig_9}, the Mott IMT (and the concomitant magnetic collapse transition) is found to be accompanied by a sudden change of the electronic state of Mn, Fe, and Co from a $3d^{N}$ to $3d^{N+1}$ configuration ($3d^{N+1}$ state turns to become favorable for Mn to Co), suggesting a crossover of the nominal oxidation state from 2+ to a nearly 1+ state (i.e, the valence of oxygen shifts near to 1-). Below the phase transition, in a correlated metal regime, the electronic state in MnO-CoO can be characterized as a mixed-valent state with a major contribution due to the $3d^{N+1}$ state (about 50\%), which has a substantial admixture of the $3d^{N}$ and excited $3d^{N+2}$ states. 
Our results therefore suggest a possible change of the oxygen valence state under high pressure which is not 2- as at low pressure but rather varies near to 1- due to oxygen-oxygen interactions. We note that similar behavior has been proposed to occur in iron oxide FeO$_2$, where an altered valence state of oxygen around to 1- instead of 2- value was found \cite{NatComm.10.153}.  
This suggests that oxygen may also have multiple valence states in oxide minerals under deep Earth conditions.

Interestingly our results for NiO also reveal a sudden change of the $3d^8$ atomic configuration weight at the Mott transition. However, in contrast to MnO-CoO, in a correlated metal phase the $3d^8$ (i.e., Ni$^{2+}$) states remain to be predominant, with a weight of above 55\% and a large $\sim$40\% admixture of the $3d^9$ excited state (below $\sim$0.47$V_0$, corresponding to $\sim$700 GPa), implying a crossover from localized to itinerant moment behavior under pressure. Our results therefore suggest the absence of the valence crossover in NiO. 
Moreover, this behavior seems to be consistent with a very slight change of the total Wannier $3d$ charge in NiO under pressure. It increases by $\sim$0.1 electrons upon compression to $\sim$400 GPa (the oxygen $2p$ charge respectively decreases). On the other hand, the change of the Wannier $3d$ charge in MnO, FeO, and CoO is more significant (up to 0.4-0.5 electrons) upon pressurizing to $\sim$150 GPa that seems to be a consequence of emptying of the antibonding $e_g^\sigma$ states under pressure. The latter leads to a different strength of covalent $p$-$d$ bonding above and below the magnetic collapse transition.

We also compute the pressure-induced evolution of the local magnetic state of paramagnetic MnO-NiO. In Figure~\ref{Fig_10} we display our results for the corresponding probabilities of the spin-states of the $3d$ electrons (e.g., in MnO, these are the HS: $|d^5_{S_z=5/2}\rangle+|d^6_{S_z=2} \rangle+|d^7_{S_z=3/2} \rangle$, IS: $|d^5_{S_z=3/2}\rangle+|d^6_{S_z=1}\rangle$, and LS: $|d^5_{S_z=1/2} \rangle+|d^6_{S_z=0}\rangle+|d^7_{S_z=1/2} \rangle$ states; contributions due to the other states are $\le$1\%) as a function of volume. At ambient pressure, our DFT+DMFT calculations reveal a high-spin magnetic state of the Mn$^{2+}$, Fe$^{2+}$, Co$^{2+}$, and Ni$^{2+}$ ions \cite{IL15,BK12,KL08}, consistent with the local magnetic moments of $\sim$1.8$\mu_B$--4.8$\mu_B$ for NiO-MnO (in a cubic crystal field, the Mn$^{2+}$ with a $3d^5$ electronic configuration and Ni$^{2+}$ $3d^8$ ions have a local moment of 5$\mu_B$ and 2$\mu_B$, respectively).

Upon metallization, the high spin-state configurations weights of MnO, FeO, and CoO exhibit a substantial drop by about 45-80\%, associated with the collapse of local magnetic moments. Most importantly, our results show that all these materials, in fact, exhibit qualitatively different microscopic behavior of their magnetic states under pressure. The magnetic collapse of MnO is associated with the appearance of the intermediate spin state (IS: $|d^5_{S_z=3/2}\rangle+|d^6_{S_z=1} \rangle$) with a weight of $\sim$45\% (mainly due to a $\sim$32\% contribution of the IS $|d^6_{S_z=1} \rangle$ states), strongly competing with the LS state ($\sim$35\%), as shown in Figure.~\ref{Fig_10}. That is, the pressure-induced correlated metallic state of MnO appears to be characterized in terms of the IS-LS fluctuations, emerging near the Mott transition. 

At the same time, the magnetic collapse in CoO is found to be driven by a substantial drop of the HS weight by $\sim$45\%, accompanied by a simultaneous increase of the LS weight (see Figure~\ref{Fig_10}). This results in a remarkable coexistence of the HS and LS states, with the HS $|d^8_{S_z=1}\rangle$ giving a major contribution of $\sim$42\%. Interestingly, we obtain that the HS state of CoO remains to be dominant up to high compression of $\sim$0.6~$V_0$ (corresponding to $\sim$190 GPa), suggesting the absence of a HS-LS crossover in CoO. This means that the magnetic collapse of CoO is associated with frustration of the HS and LS states rather than with the HS-LS transition. 

In contrast to that FeO shows a substantial drop of the HS state configuration weight down to $\sim$20\% and concomitant population of the LS state to about 70\% under pressure. The latter originates from a quantum superposition $\sqrt{0.22} |d^6_{S_z=0}\rangle + \sqrt{0.42} |d^7_{S_z=1}\rangle$. This suggests that FeO undergoes a conventional HS-LS crossover, which is possibly complicated by the presence of a valence crossover. Moreover, it seems that the HS-LS transition appears slightly below the Mott IMT in FeO, at $\sim$0.67~V$_0$, suggesting that the Mott and spin-state transitions are decoupled at high temperature \cite{Greenberg}. Considering NiO at high compression to 0.5~$V_0$ ($\sim$500 GPa), below the Mott IMT, we observe that a reduction of the HS configuration is insufficient, $\leq$15\%, implying that the Mott transition does not alter the HS state of NiO. 
Overall, our results suggest the emergence of quantum critical valence (charge) and spin-state (spin) fluctuations near the pressure-driven Mott IMT in these compounds (correlated systems with spin-state active ions), which may have important implications for the understanding of quantum criticality of the Mott transitions \cite{q_criticality_exp,q_criticality_teor}.

\section*{V. CONCLUSION}

In conclusion, we determine the electronic structure, magnetic state, and structural properties
of correlated transition metal monoxides MnO, FeO, CoO, and NiO in the rocksalt B1 crystal structure at high pressure and temperature using the DFT+DMFT method. We obtain that under pressure MnO-NiO exhibit a Mott IMT which is accompanied by a simultaneous collapse
of local magnetic moments and lattice volume, implying a complex interplay between chemical bonding and
electronic correlations. We explain a monotonous decrease of the Mott IMT transition pressure $p_c$ which varies from $\sim$145
to 40 GPa, upon moving from MnO to CoO, and then suddenly increases to $\sim$429 GPa. We provide a unified picture of such a behavior and suggest that it is
primarily a localized to itinerant moment behavior transition at the Mott IMT that gives rise to magnetic collapse in transition metal oxides.

We have shown that the interplay between electronic correlations, spin state, and the lattice in the vicinity of the Mott IMT results in the formation of a complex electronic and magnetic states of MnO, FeO, CoO, and NiO under pressure. 
In particular, the Mott IMT and collapse of the local moments in MnO, FeO, and CoO under pressure are found to be accompanied by a sharp crossover of the valence state, implying a complex interplay between the charge, spin, and lattice degrees of freedom. 
This suggests a remarkable importance of the valence fluctuations for understanding the electronic state of correlated systems near the Mott transition. In fact, this is in connection to quantum critical phenomena, i.e., the quantum critical nature of the Mott transition and (possible) quantum valence criticality near the Mott transitions. 
Most importantly, we provide a novel microscopic explanation of the magnetic collapse of all these compounds. On the basis of our DFT+DMFT calculations we observe three distinct scenarios: a) magnetic collapse caused by the appearance of the IS state, strongly competing with the LS state in MnO; b) a conventional HS-LS crossover in FeO, and c) a remarkable coexistence of the HS and LS states (HS-LS frustration) in CoO.
We propose that quantum fluctuations of the valence and spin states emerging near the Mott transition may have important implications for the understanding of quantum criticality of the Mott transitions.
Finally, we point out the importance of further theoretical and experimental investigations of the above discussed correlated compounds using, e.g., x-ray absorption spectroscopy and electron energy-loss spectroscopy, which are powerful probes of a valence state.

\begin{acknowledgments}
We thank L. Pourovskii, A. Georges, R. Jeanloz, G. Kh. Rozenberg, and D. I. Khomskii for valuable discussions. 
Theoretical analysis of the magnetic and valence states as well as of the structural properties of MnO, FeO, CoO, and NiO was supported by the Russian Science Foundation (project No. 18-12-00492).
Simulations of the electronic structure were supported by the state assignment of Minobrnauki of Russia (theme ``Electron'' No. AAAA-A18-118020190098-5).
Support from Knut and Alice Wallenberg Foundation (Wallenberg Scholar Grant No. KAW-2018.0194), the Swedish Government Strategic Research Area in Materials Science on Functional Materials at Link\" oping University (Faculty Grant SFOMatLiU No. 2009 00971), and the Swedish e-Science Research Centre (SeRC) is gratefully acknowledged.

\end{acknowledgments}

\end{document}